

\magnification=\magstep1
\baselineskip=24 true pt
\hsize 15. truecm
\vsize 21. truecm
\bigskip
\bigskip
\centerline {\bf THEORY OF LOOPS AND STRINGS WITH MATTER}
\centerline {\bf IN THE ADJOINT REPRESENTATION}
\vfill
\centerline {\bf  JNANADEVA MAHARANA }
\bigskip
\centerline {\it { Institute of Physics, Bhubaneswar-751005, INDIA}}
\bigskip
\centerline {and}
\bigskip

\centerline {\bf LAMBODAR P. SINGH }
\bigskip
\centerline {\it { Department of Physics, Utkal University,
Bhubaneswar-751004, INDIA}}
\vfill
\centerline {\bf Abstract}

  We have presented canonical and path integral formulations of a theory
of loops and closed strings with the matter field quanta transforming
in the adjoint representation of the SU(N) gauge group. The physical
processes arising out of the interactions of loops and closed strings
are discussed.
\bigskip
\bigskip

PACS Index:$ 11.17+Y$
\vfill
\eject
\noindent {\bf {I.INTRODUCTION.}}

  It is now accepted that the principle of local gauge invariance
determines all fundamental interactions. The standard model
consisting of quantum chromodynamics (QCD) and the electroweak
theory built on such gauge principles have been vindicated by the
experiments to a great degree of accuracy. One of the intriguing
feature of QCD is
that the observed particle spectrum belong to the coloured singlet
sector. Thus the machanism of colour confinement presents itself as a
challenging problem to contend with. As an alternative to building a
theory with coloured gauge and matter fields, it has been argued that
the theory should be constructed in terms of gauge invariant
observables to startwith. Mandelstam [1] had pioneered this approach
long ago and subsequently Polyakov [2] has emphasised the importance
of the construction. In early eighties Makeenko and Migdal [3]
presented a loop-space formulation of QCD and derived the well known
loop equation. One of the important ingredient of Makeenko-Migdal
formulation is the $1/N$ expansion expounded beautifully in a seminal
paper by 't Hooft [4] in the context of two-dimensional QCD. It is well
known that $1/N$ expansion , in this context, is also a topological
expansion which brings out the nonperturbative aspect of the theory
in an elegant manner.

  The simplest scenerio is to consider the pure $SU(N)$ Yang-Mills
theory where the Wilson loop operator $W(L)$,

 $$   W(L)=Tr P exp  [i \oint A_{\mu} d\zeta^{\mu}] \eqno (1) $$

\noindent is the gauge invariant observable. The fields $A_\mu$ are
lie algebra valued; $A{\mu} =A{\mu}^a T^a$ where $T^a$ are the
generators of the gauge group . Here `Tr' denotes the trace over the
$SU(N)$ color indices and P denotes path the ordering. In the
presence of the matter fields belonging to the fundamental
representation, there is an open string with matter sources at each
end connected by coloured flux tubes. Such objects can be identified
with "mesons". Other coloured singlet objects too can be constructed
within the scope of the theory.
  Recently there is renewed interest in the loop equations due to
their importance in $2-$ dimensional gravity and matrix models [5].
Recently,  Dalley and Klebanov [6] have studied the dynamics of gauge fields
interacting with matter fields belonging to the adjoint
representstion of the gauge group in $1/N$ aproximation in two
space-time dimensions. One of the
interesting features of this model is the existence of the closed
string states consisting of matter fields besides the Wilson loop
operator. The gauge invariant string operator we work with are
defined as,

$$    M(S)= Tr \phi (x) U(x,y) \phi (y) U(y,x)  \eqno (2) $$
\noindent where  $$ U(x,y) = P exp[i\int A_\mu d\zeta^\mu].\eqno(3)$$
\noindent The dynamics of this theory when formulated in terms of
closed strings $M(S)$  and the Wilson loop $W(L)$ operators is quite
interesting.

 The purpose of this article is to present a systematic study of the
dynamics of such types of loops and strings. First we present a
canonical formulation of the theory and subsequently develope a path
integral formulation in the phase space of gauge invariant variables
mentioned above. It is worthwhile to point out that a canonical
formulation in the context of Wilson loop operator was carried out
earlier and later extended to include spinless matter fields in the
fundamental representation [7]. The present work is based on our
previous path integral formulation of such a theory where large-N
limit of QCD was studied.

\noindent {\bf {II.CANONICAL FORMALISM FOR LOOPS AND STRINGS :}}

  We start with the Hamiltonian in the Phase-space of Yang-Mills and
matter fields in two dimensions. In the $A_0 = 0$ gauge, this has the form,

$$ H = \int dx \bigg[ {1\over 2} \pi^a_{k}(x)\pi^a_{k}(x) + {1\over 4} F^a_{ij}
F^a_{ij}(x) + \pi^a_{\phi}(x) \pi^a_{\phi}(x)$$$$ +
D_k\phi^{a}(x)D_k\phi^{a}(x) +
\mu^2
\phi{^a}(x)\phi{^a}(x) + {\lambda \over 2}
(\phi^a(x)\phi^a(x))^2\bigg]\eqno(4)$$

where $$  D_k\phi^a(x) = \partial_k\phi^a(x) + i g f_{abc}A^b_i\phi^{c}
\eqno(5)$$

and $$ F^a_{ij}(x) = \partial_iA^a_j(x) -\partial_jA^a_i(x)
+ i g f_{abc} A^b_i A^c_j.\eqno(6)$$
\noindent $f_{abc}$ are the structure constants of the gauge hroup
given by

 $$      [ T_a , T_b ] = f_{abc} T_c. \eqno(7)$$

\noindent The generators T's satisfy the relation

$$    ( T_a)_{\alpha\beta} ( T_a)_{\rho\sigma} = \delta_{\alpha\sigma}
\delta_{\beta\rho} -{1\over N}\delta_{\alpha\beta} \delta_{\rho\sigma}
\eqno(8)$$

\noindent $\pi^a_k(x)$ and $\pi^a_{\phi}(x)$ occurring in eq.(4) are the
canonical momenta
conjugate to $A^a_k(x)$ and $\phi^a(x)$ respectively satisfying the usual
commutation relations,

$$\eqalign{[ \phi^a(x),\pi^b_{\phi}(y)] = i \delta_{ab}
\delta(x-y)\cr  [ A^a_i(x), \pi^b_i(y)] = i \delta_{ab}
\delta(x-y)\cr}\eqno(9)$$

\noindent and all other commutation relations vanish.

  Choosing to work in the Schr$\ddot{o}$dinger representation, one defines the
wave fuction $\Psi(\vec A,\phi)$ as a functional of $\vec A$ and $\phi$
satisfying

$$         i{\partial\over {\partial t}}\Psi = H\Psi \eqno(10)$$

\noindent In this representation, the canonical momenta can be written as,

$$    \pi^a_i(A) = - i{ \delta\over {\delta A^a_i(x)}}\eqno(11a)$$
and
$$    \pi_{ \phi}(x) = -i {\delta\over {\delta\phi(x)}}\eqno(11b)  $$

so that Hamiltonian can be expressed in the following form

$$ H = \int dx\bigg[ -  {1\over 2}{({\delta \over {\delta {A^a}_i(x)}})^2}
+{1\over 4}(F^a_{ij})^2 + $$

$$\bigg\{ -({\delta\over {\delta\phi^a(x)}})^2
+ D_k\phi^{a}(x) D_k\phi^{a}(x) + \mu^2{\phi^{a}(x)}^2 + {\lambda\over 2}
(\phi^a\phi^a)^2\bigg\} \bigg] \eqno(12)  $$

  	In order to recast the above theory in terms of closed loops W(L)
and closed strings M(S), the wave fuction $\Psi(\vec A,\phi)$ is rewritten
as a functional of W(L) and M(S);

$$            \Psi(\vec A,\phi) = \Psi(W,M)\eqno(13)  $$

\noindent The transformation from the space of $ \vec A$ and $\phi$ to
W and M is
assumed to be a canonical one as stated above. Thus, correspondingly
the Hamiltonian can be wrutten in terms of W,M and their conjugate
momenta $\pi_w(\tilde L)$ and $\pi_M(\tilde S)$ where $\tilde L$ and
$\tilde S$ are the conjugate operators with paths traversed in the
opposite direction to L and S respectively. These conjugate momenta
satisfy the commutation relation,

$$  [ W(L_1),\Pi_W({\tilde L}_2)] = i\delta (L_1,{\tilde L}_2)\eqno(14a)$$
and
$$  [ M(S_1), \Pi_M({\tilde S}_2)]= i\delta (S_1,{\tilde S}_2)\eqno(14b)$$

\noindent where the delta functions $\delta (L_1,{\tilde L}_2)$ and $\delta
(S_1,
{\tilde S}_2)$ are defined in the loop and string space. The
conjugate momenta  can be represented by the functional derivative,

$$ \Pi_W( \tilde L) = -i {\delta\over {\delta W(\tilde L)}}\eqno(15a)$$

and

$$ \Pi_M(\tilde S) = -i{\delta\over {\delta M(\tilde S)}} \eqno(15b) $$

  The differential operators occurring in the kinetic energy operators
can be converted to the ones involving the new dynamical variables
W(L) and M(S) using the chain rule of differentiation to obtain the
kinetic part of the Hamiltonian in the form [7],

$$ H_{kin} = \bigg[ i\sum_L d_W(L)\Pi_W(\tilde L) + \sum_{L_1,L_2} D_W
(L_1,L_2)\Pi_W(\tilde L_1)\Pi_W(\tilde L_2)$$$$ + i\sum_S d_M(S)\Pi
_M(\tilde S) + \sum_{S_1,S_2} D_M(S_1,S_2)\Pi_M(\tilde S_1)\Pi_M
(\tilde S_2)$$$$  + \sum_{L,S} D_{WM}(L,S) \pi_W(\tilde L)\pi_M(\tilde S)
\bigg] \eqno(16)  $$

\noindent where the d's and D's are the polynomials of loop and string fields
as
given below.

$$ d_W(L) = -\int dx {\delta^2\over {\delta {A^a_i(x)}^2}} W(L)$$

$$\quad     = \oint dx \oint dx'\delta( x -  x')
 W(L) N(1-{1\over N}) $$$$- \sum_{{L_1}', {L_2}'} 2\delta
(L_1, {L_1}')
\delta (L_2, {L_2}')W({L_1}') W( {L_2}') \eqno(17) $$

$$ D_W(L_1,L_2) = \int dx {\delta W(L_1)\over {\delta A_i^a(x)}}
{\delta W(L_2)\over {\delta A^a_i(x)}}$$$$
                = N\oint_{L_1}dx \oint_{L_2}dx' \delta
(x -  x')
\delta (L_1,\tilde L_2) - \sum_{L3} \delta (L_1+L_2,L_3) W(-L_3)$$$$
- {1\over N} {\oint}_{L_1}dx {\oint}_{L_2}dx'\delta (x -  x') W(L_1)
W(L_2) \eqno(18) $$

$$ d_M(S) = - \int dx {\delta^2 M(S)\over {\delta{\phi(x)}^2}} - \int
dx {\delta^2 M(S)\over {\delta {A^a_i(x)}^2}}$$$$ = -2\sum_{L_1,L_2}
\delta (S,L_1+L_2) W(L_1) W(L_2) - \int_S dx\int_Sdx'
\delta ( x - x')M(S)(N-{4\over N})$$$$ - \sum_{S',L'}\delta (L,S'+L')
W(L') M(S') - 2\sum_{S_1(1),S_2(1)}\delta (S,S_1(1)+S_2(1))
M(S_1(1)) M(S_2(1)) \eqno(19) $$

$$ D_M(S_1.S_2) = \int dx {\delta M(S_1)\over {\delta A^a_i(x)}}
{\delta M(S_2)\over {\delta A^a_i(x)}} + \int dx {\delta M(S_1)\over
{\delta \phi(x)}} {\delta M(S_2)\over {\delta \phi(x)}}$$$$ = 2\sum_{S(4)}
\delta (S_1+S_2,S(4)) M(S(4)) -{1\over N} \int dz\int dz'
\delta (z -  z') M(S_1) M(S_2) $$$$+ \sum_{z=x_1x_2,x_1y_2,
y_1x_2,y_1y_2} \delta (S_1+S_2,S(z)) M(S)\eqno(20)  $$

\noindent and

$$ D_{WM}(L,S) = \int dx {\delta W(L)\over {\delta A^a_i(x)}}
{\delta M(S)\over {\delta A^a_i(x)}} = 2\sum_{S'}\delta (S',L+S) M(S')
$$$$-{1\over N}\int dz\bigg[\int_{x'}^{y'} dz'\delta ( z -
z') W(L) M(S) + \int_{y'}^{x'} d{\vec z}'\delta (\vec z-
{\vec z}') W(L) M(S)\bigg]\eqno(21) $$

  The coefficients $d_W(L),D_W(L_1,L_2),d_M(S),D_M(S_1,S_2),D_{WM}
(L,S)$ appearing in the expression (16) for Hamiltonian and defined
by eqs. (17)- (21) describe various physical processes as follows.
The $d_W(L)$ and $D_W(L_1,L_2)$ describe the same processes as in
the case of matter fields in the fundamental representation since
W(L) operator retains the same form as in earlier works. For the
sake of completeness we mention that the first term of $d_W(L)$
describes a loop remaining a loop and the second term describes a
loop splitting into two loops $L_1$  and $ L_2$ by pinching (Fig.1a).
The first term of $D_W(L_1,L_2)$ denotes annihilation of two loops
$L_1$ and  $L_2$ in the case of $L_2$ being conjugate to $L_1$ (Fig.
1b) and the second term stands for  two loops $L_1$ and$L_2$ fusing
to form a single loop $L_1+L_2$ (Fig.1c). The third term describes
two loops remaining as two loops.

  The first term of $d_M(S)$ denotes a string $S$ remaining the same,
the second term describes splitting of $S$ to two loops $L_1$ and$L_2$
(Fig.2a). The third term describes the string splitting to form a loop
 and a string (Fig.2b). The last term stands for splitting of a string
to two strings with one matter field each  $ S(1)$ (Fig 2c).The terms in the
expression
for $D_M(S_1,S_2)$ have the physical interpretation as follows; the
first term denotes two types of fusion of two strings to form a string
with four matter fields $S(4)$ (Fig. 3a), the second term denotes the two
strings
remaining as two strings and the last term gives again the fusion of
two strings to form one string but with two matter fields in the form
of a figure-of-eight (Fig.3b). The first term in the expression for
$D_{WM}(L,S)$ describes the fusion of the string $S$ and the loop $L$ to
form a string $S'$ (Fig.4) whereas the second term stands for loop
and string remaining as such.

  Here again one finds that in case of all two-string operators like
$D_W(L_1,L_2)$, $D_M(S_1,S_2)$ and $D_{WM}(L,S)$
the possibility of two objects remaining the same two objects is
suppressed by a factor of ${1\over N}$ with respect to other nontrivial
interactions and vanishes for U(N) group.

There are many interesting features of this model as is evident from
the discussions above. We notice that there are several new type of
loop-string processes as compared to the case when the matter fields
belong to the fundamental representation of the group e.g. the interaction
of strings with two matter fields to produce stings with one or four
matter fields $S(1)$ or $S(4)$ or with two matter fields in the shape
of a figure-of-eight.

  The potential energy terms in the Hamiltonian (Eq. 12) can be expressed
in terms of loop and string operators defined on infinitesimal loops and
strings. For infinitesimal loop operator, as shown earlier [7], the loop
operator can be expressed as

$$ W(\sigma (x)) = Tr P exp (i \oint  Adx )$$$$
                 = Tr exp [i {{F_{ij}}^a} \sigma_{ij}]$$$$
                 = N - {1\over 2} (F_{ij}^a \sigma_{ij})^2\eqno(22)  $$
leading to

$$      {\delta ^2 W(\sigma (x))\over {(\delta \sigma_{ij})^2}} =
- (F_{ij}^a(x))^2   \eqno(23)    $$

Thus the potential energy term $\int dx (F_{ij}^a(x))^2$ can be written
as

$$   \int dx (F_{ij}^a(x))^2 = \sum_L\tau(L) W(L) \eqno(24) $$
with
$$ \tau(L) = \int dx {\delta^2\over {(\delta \sigma_{ij})^2}}
\delta (L,\sigma(x)) \eqno(25)  $$

  To express the rest of potential energy terms in the Hamiltonian in
terms of loops and strings, we note that,

$$   \bigtriangledown_x M(s) = Tr {D_x\phi(x)}U(x,y)\phi(y)
U(y,x) \eqno(26a) $$
and
$$  \bigtriangledown_y M(S) = Tr \phi(x) U(x,y){D_y\phi(y)} U(y,x)
\eqno(26b) $$

These relations are closed string processes involving two matter field
quanta and two string operator U(x,y) and U(y,x). This is to be contrasted
with the earlier case where two matter fields in fundamental representation
and a single string operator U was involved. Eq. 26 leads to,

$$  Tr (D_k\phi) (D_k\phi) = \lim_{y\rightarrow x}  \bigtriangledown_y
 \bigtriangledown_x M(S) \eqno(27) $$

\noindent where the limit $y\rightarrow x$ is taken along the paths to
make the
string an infinitesimal one. The potential energy terms can be formally cu
tu

into the same form as in the open string case and are expressed as

$$ H_{pot} = \sum_L \tau(L)W(L) + \sum_S \tau(S)M(S) + \mu^2 \sum_S
\rho(S) M(S) + {\lambda \u
er 2} \sum_S \rho(S) M(S)^{2} \eqno(28) $$

\noindent where

$$ \tau(S) = \int dx\int dy [ \bigtriangledown_x
\bigtriangledown_y \delta ( x -  y)]\delta (S,s)\eqno(29)$$
and
$$\rho (S) = \int dx\int dy \delta ( x -  y)\delta (S.s)
\eqno(30)  $$

  Now the total Hamiltonian is obtained by adding eqs.(16) and (28).
However, the Hamiltonian so obtained is not manifestly hermitian
as a consequence of the change of the Hilbert space from the space of
field operators $\phi(x)$ and $\vec A (x)$ to that of loops and strings.
This problem can however be tackled by a similarity transformation as
has been worked out in detail in Ref.7.

\noindent {\bf {III. THE PATH INTEGRAL FORMULATION :}}

  We now wish to present a path integral formulation of the theory
discussed above.

  In order to  develope the path integral formulation, we start
with the SU(N) invariant Lagrangian density of the theory involving the
gauge fields $A_\mu(x)$ ($A_0 \not=0$) and the scalar matter field $\phi
(x)$ wtitten as,

$$ {\cal L} = Tr [ - {1\over 4}F_{\mu\nu}F^{\mu\nu} + D_{\mu}\phi
D^{\mu}\phi - {\mu}^2 {\phi}^2 - {\lambda \over 2}{\phi}^4 ]\eqno(31)  $$

where $D_{\mu}\phi$ and $F_{\mu\nu}$ have the component decomposition as
given by eqs.(5) and (6) and can be put as

$$ D_{\mu}\phi = \partial_{\mu}\phi + i [A_{\mu},\phi]  \eqno(32a)  $$
and
$$ F_{\alpha,\beta} = \partial_{\alpha} A_{\beta} - \partial_{\beta} A_{\alpha}
+ i[ A_{\alpha} , A_{\beta} ]   \eqno(32b) $$

  As is well known this theory possesses constraints which can easily
be seen by looking at the canonical conjugate momenta with respect to
the gauge and the matter fields. The canonical conjugate momenta are,

$$    {\Pi_{\mu}}^a(x) \equiv {\partial {\cal L} \over {\partial{\dot A}_\mu^
a(x)}} = F_{0\mu}^a(x)  \eqno(33)  $$
and
$$    \Pi_{\phi}^a(x) \equiv {\partial {\cal L}\over {\partial {\dot \phi}^a
(x)}} = D_0 {\phi}^{a}(x)  \eqno(34)  $$

It thus follows that

$$          \Pi_0^a(x) = 0 \eqno(35)  $$

\noindent are a set of primary constraints [9]. The canonical Hamiltonian is
obtained as,

$$ {\cal H}_c(x) = {1\over 4} F_{ij}^a(x) F_{ij}^a(x) + {1\over 2} \Pi_k^a(x)
\Pi_k^a(x) - A_0^a(x)D_k\Pi_k^a(x) $$$$+ \Pi_\phi^a(x)\Pi_\phi^a(x) -
i g f_{abc}\Pi_\phi^{a}(x) A_0^b(x) \phi^c(x)$$$$ + D_k\phi^u
x)D_k\phi^a(x)
+ \mu^2 \phi^a(x)\phi^a(x) + {\lambda\over 2} (\phi^a(x)\phi^a(x))^2
\eqno(36)  $$

\noindent This expression agrees with eq.(4) for $A_0^a = 0$. The total
Hamiltonian
du
su
y can now be written as,

$$  {\cal H}_T(x) ={\cal H}_c + \sum_{a=1}^{N^2-1}{\cal u}^a\Pi_0^a
\eqno(37) $$

In order that the primary constraint remains preserved for all times
its Poission bracket with the total Hamiltonian must vanish and that
leads to the secondary constraint,

$$     F^a = 0   \eqno(38)  $$

where, $$ F^a(x) = G^a(x) + K^a(x)  \eqno(39)  $$
with   $$ G^a (x) = D_k\Pi_k^a(x) \eqno(40)   $$
and $$        K^a(x) = i g f_{abc} \Pi^b_\phi(x) \phi^c(x) \eqno(41) $$

The $F^a$'s  do not generate any new constraints. Thus the partition
function in the $[A^a_\mu(x), \Pi^a_\mu(x), \phi^a(x), \Pi_\phi^a(x)]$
phase space can be written as,

$$ {\cal Z} = \int {\cal D}(A_\mu^a){\cal D}(\Pi_\mu^a){\cal D}(\Pi_\phi^a)
{\cal D}(\phi^a) \delta (\Pi_0^a) \delta (F^a) $$$$exp\; [i (\Pi_\mu^a\dot A
_\mu + \Pi_\phi^a\dot \phi^a - {\cal H}_T ) d^3x ] \eqno(42) $$

  We now wish to go over to the phase-space of loops and strings
defined earlier. To begin with we replace the commutation relations
( eq. 9) by the ones between the gauge invariant operators and their
conjugate momenta (eq. 14) as dicussed earlier. In this approach the
loop and string operators are treated as a complete set of independent
variables. This independence, as we know, holds good strictly in the
limit of $N \rightarrow \infty$ [7]. The transformation from the space
of gauge and matter fields to that of loops and strings is again assumed
to be canonical one. However, in developing the path integral formulation
we do not commit to any specific representaion of the momentum operators
conjugate to loops and strings.We obtain expressions for them invoking
the invariance of the Poission and Langrange brackets under a canonical
transformation instead, as [8]

$$ \Pi^a_i(x) = \sum_L {\delta W(L)\over {\delta A^a_i(x)}} \Pi_W(L)
+ \sum_S{\delta M(S)\over {\delta A^a_i(x)}}\Pi_M(S) \eqno(43) $$
and
$$  \Pi_\phi^a(x) = \sum_S{\delta M(S)\over {\delta \phi^a(x)}} \Pi_M(S)
\eqno(44)  $$

Therefore, the K.E. part of the gaugefields can be written as,

$$ \Pi^a_i(x) \Pi^a_i(x) = \bigg[\sum_L {\delta W(L)\over {\delta A^a_i(x)}}
\Pi_W(L) + \sum_S{\delta M(S)\over {\delta A^a_i(x)}} \Pi_M(S)\bigg]\times
$$$$\bigg[\sum_{L'}{\delta W(L')\over {\delta A^a_i(x)}} \Pi_W(L') + \sum_{S'}
{\delta M(S')\over {\delta A^a_i(x)}} \Pi_M(s')\bigg] \eqno(45) $$

  This expression exactly agrees with the gauge field part of the
eq.(16) if one adopts the specific representation of the form given
by eq.(15). The matter field kinetic energy terms similarly can be
expressed in terms of gauge invariant operators using eq.(43). The
potential energy too can be expressed in terms of loops and strings,
as described before, in the form of eq.(28). The Euclidean Hamiltonian
can now be expressed as,

$$ H = {1\over 2} \sum_{L,L'} g^{LL'}\Pi_W(L)\Pi_W(L') + \sum_{L,S}
g^{LS}\Pi_W(L)\Pi_M(S) +{1\over 2}\sum_{S,S'}g^{SS'}\Pi_M(S)\Pi_M(S
')$$$$ + \sum_{S,S'} G^{SS'}\Pi_M(S)\Pi_M(S') + \sum_L\tau(L)W(L)
+\sum_S\tau(S)M(S) $$$$ + \mu^2\sum_S\rho(S)M(S) + {\lambda\over 2}
\sum_S\rho(S)M(S)^2 \eqno(46)  $$

where the various coefficients are Euclidean analogs of the coefficients
appearing in eq.(16) and (28) and have the form,

$$ g^{LL'} = -\int d^2x {\delta W(L)\over {\delta A^a_\mu(x)}}
{\delta W(L')\over {\delta A^a_\mu(x)}} \eqno(47)  $$
$$ g^{LS} = -\int d^2x {\delta W(L)\over {\delta A_\mu^a(x)}}{\delta M(S)
\over {\delta A_\mu^a(x)}} \eqno(48)  $$
$$  g^{SS'} = -{1\over 2}\int d^2x {\delta M(S)\over {\delta A_\mu^a(x)}}
{\delta M(S')\over {\delta A_\mu^a(x)}} \eqno(49)  $$
$$  G^{SS'} = \int d^2x {\delta M(S)\over {\delta \phi^a(x)}}{\delta M(s')
\over {\delta \phi^a(x)}}  \eqno(50)  $$
$$  \sigma(L) = -{1\over 4} \int d^2x {\delta^2\over {(\delta \sigma_ij)^2}}
\delta (\sigma (x),L)  \eqno(51)  $$
$$  \tau(S) = \int d^2x d^2y \partial_x\partial_y \delta^2(x-y)\delta
(s,S)  \eqno(52) $$
$$  \rho(S) = \int d^2x d^2y \delta^2(x-y) \delta (s,S)  \eqno(53)  $$

  To obtain an expression for the partition function Z in the phase space
 of gauge invariant fuctionals, as is well known, it is not sufficient
to replace eq.(46) in the exponent of eq.(42). There is an additional
term $\bigtriangleup V$ and this can formally be obtained in a manner
analogous to the procedure followed earlier. Thus the partition function
can be put in the form,

$$   Z = \int [g(W)]^{-{1\over 2}} [G(M)]^{-{1\over 2}}{\cal D}[W(L)]
{\cal D}[M(S)]{\cal D}[\Pi_W(L)]{\cal D}[\Pi_M(S)] $$$$
exp\bigg[ - \bigg\{ {1\over 2} \sum_{L.L'} g^{LL'}\Pi_W(L)\Pi_W(L')
+ \sum_{L,S} g^{LS}\Pi_W(L)\Pi_M(S)$$$$ + {1\over 2} \sum_{S,S'}
g^{SS'}\Pi_M(S)\Pi_M(S')  + \sum_{S,S'} G^{SS'}\Pi_M(S)
\Pi_M(S') + \sum_L\tau(L)W(L)$$$$ + \sum_S [\tau(S)+\mu^2\rho(S)]
M(S) +{\lambda\over 2} \sum_S\rho(S)M(S)^2 + \bigtriangleup V
\bigg\}\bigg]  \eqno(54)   $$

where g(W) and G(M) are the determinants of the metric tensors
$g_{LL'}$ and $G_{SS'}$ which, in turn, are inverse of the tensors
$g^{LL'}$ and $G^{SS'}$ respectively. The $\Pi_W(L)$ and $\Pi_M(S)$
integrations can be performed using the standard procedure enabling
in principle to compute the expectation value of any function of loop
and string variable using the above partition function.

  We note that the coefficients $g^{LL'}$,$g^{LS}$,$g^{SS'}$ and $G^{SS'}$
describe the dynamics of loops and the closed strings as discussed earlier
being related to the operators $D_W(L_1,L_2)$,$D_{MS}(L,S)$ and
$D_M(S_1,S_2)$ respectively.

\noindent {\bf {IV. DISCUSSIONS : }}

  In this article we have developed both canonical and path integral
formulations of a SU(N) gauge theory with scalar matter field in
the adjoint representation of the gauge group. This is the closed string
analog of the earlier works [7,8]. In such a formalism we stress that
any number of matter quanta can appear in the closed string operator
[6]. We have confined ourselves to the closed string operators defined
with two matter fields and have interpreted the various terms occurring
in the Hamiltonian in terms of the interactions between loops and such
strings mainly for the sake of comparision with the scenerio enunciated
earlier [7,8].

  The usefulness of the path integral formulation lies in the speculation
that given the action in terms of loops and strings the generating
functional can be computed enabling, in turn, to develope a perturbation
theory in analogy with quantum field theories and compute n-point functions
in loop and string space.

  We mention at the end that our work can be generalised to the case of
fermionic matter fields in the light of the work of Dalley nad Klebanov
[6] and to a supersymmetric formulation as well.

\centerline {\bf { ACKNOWLEDGEMENTS:}}

\centerline {\bf {APPENDIX :}}

  We now give the derivation of the coefficients appearing in the
expression for the total Hamiltonian insofar as the ones which are
different from the earlier work. Thye difference arises because of a
different field structure of the closed string configuration we work with,
which in turn is a consequence of the matter field being in the adjoint
representation of the gauge group.

  We start with $d_M(S)$. First we calculate $\int dx {\delta ^2M(S) \over
{ (\delta A^{a}_i)^2}}$. With the definition of $M(S)$ we have, we get,

$$\int dx{\delta^2M(S) \over { (\delta A^{a}_i)^2}} = $$$$\int^x_y dz
\int^y_xdz'\delta (z-z')\phi_{\alpha \beta }(x) [ T_a (z)U(x,y)]_
{\beta \gamma} \phi_{\gamma \sigma}(y) [ T_a(z') U(y,x)]_{\sigma \alpha}
$$$$+ \int_x^ydz\int_y^xdz'\delta (z-z')\phi(x)_{\alpha\beta}[T_a(z')
U(x,y)]_{\beta\gamma}\phi(y)_{\gamma\sigma}[T_a(z)U(y,x)]_{\sigma\alpha}$$$$
 + \int_y^x dz\int_y^x dz'\delta (z-z')\phi(x)_{\alpha\beta}[T_a(z)T_a(z')
U(x,y)]_{\beta\gamma}\phi(y)_{\gamma\sigma}U(y,x)_{\sigma\alpha} $$$$
+ \int_ x^y dz\int_x^y dz'\delta (z-z') \phi(x)_{\alpha\beta}U(x,y)_
{\beta\gamma} \phi(y)_{\gamma\sigma}[T_a(z)T_a(z')U(y,x)]_{\sigma\alpha}
\eqno(A.1)$$

  In the first two terms z and $z'$ can not be the same point on the
string since z occurs in the $x \rightarrow y$ sector and $z'$ occurs in
the $y\rightarrow x$ sector. Looking at the first term first, we write
it as,

$$ \int_y^x dz\int_x^y dz'\delta (z-z') \phi(x)_{\alpha\beta}U(x,z)_{\beta
 p} [T_a(z)]_{pq}U(z,y)_{qr}$$$$ \phi(y)_{r\sigma}U(y,z')_{\sigma m}
[T_a(z')]_{mn}U(z',x)_{n,\alpha}  \eqno(A.2) $$

Using eq.(8), this becomes

$$ = \int_y^x dz\int_x^y dz' \delta (z-z') \phi(x)_{\alpha\beta}U(x,z)_
{\beta p}[\delta _{pn}\delta _{qm} -{1\over N}\delta _{pq}\delta _{mn}]
$$$$U(z,y)_{qr}\phi(y)_{r\sigma}U(y,z')_{\sigma m}U(z',x)_{n\alpha}
$$$$
= \int_y^x dz\int^y_x dz'\delta (z-z') \phi(x)_{\alpha\beta}U(x,z)_{\beta n}
U(z',x)_{n\alpha}U(z,y)_{qr}\phi(y)_{r\sigma}U(y,z')_{\sigma q}$$$$
-{1\over N}\int^x_y dz\int^y_x dz' \delta (z-z') \phi(x)_{\alpha\beta}
U(x,z)_{\beta q}U(z,y)_{qr}\phi(y)_{r\sigma}U(y,z')_{\sigma m}U(z',x)_
{m\alpha}\eqno(A.3)   $$

  Thus we arrive at the result where the first term represents the string
with two matter fields splitting to two strings with one matter field each
(Fig.2c) and the second term stands for the string remaining as such.

  The second term of eq(A.1) also has exactly similar contribution. The
third and the fourth terms are different because the points z and $z'$
can be both same and different points on the path because of path
ordering.

  If z and $z'$ are different points on the path, the third term of eq(A.1)
can be written as

$$ \int_y^x dz\int_y^x dz'\delta (z-z')\phi(x)_{\alpha\beta}U(x,z)_{\beta g}
[T_a(z)]_{gh}U(z,x_0)_{hi}$$$$ U(x_0,z')_{ij}[T_a(z')]_{jk}U(z',y)
_{k\gamma}\phi(y)_{\gamma\sigma}U(y,x)_{\sigma\alpha} $$$$
= \int_y^xdz \int^x_ydz'\delta (z-z') \bigg[\phi(x)_{\alpha\beta}U(x,z)
_{\beta k}U(z',y)_{k\gamma}\phi(y)_{\gamma\sigma}U(y,x)_{\sigma\alpha}
U(z,x_0)_{ji} $$$$ U(x_0,z')_{ij} - {1\over N} \phi(x)_{\alpha\beta}U(x,z)_
{\beta h}U(z,x_0)_{hi}U(x_0,z')_{ik}U(z',y)_{k\gamma}\phi(y)_{\gamma
\sigma}U(y,x)_{\sigma\alpha}\bigg]  \eqno(A.4) $$

\noindent using eq.(8).

  The first term depicts a process where the string splits to a string
and a loop (Fig.2b ) and the second term describes a string remaining
a string.

  If z and $z'$ happen to be same point on the path, the expression can
be written as,

$$ \int_y^x dz\int_y^x dz'\delta  (z-z') \phi(x)_{\alpha\beta}U(x,z)_
{\beta g}[T_a(z)]_{gh}[T_a(z')]_{hi}U(z',y)_{i\gamma}
\phi(y)_{\gamma\sigma}U(y,x)_{\sigma,\alpha} $$

Using eq.(8) again, we get

$$\int_y^xdz\int_y^xdz'\delta (z-z') \bigg[(N - {1\over N})\phi(x)_
{\alpha\beta}
U(x,z)_{\beta i}U(z',y)_{i\gamma}\phi(y)_{\gamma\sigma}U(y,x)_{\sigma
\alpha}\bigg] $$$$
= \int_y^x dz \int_y^x dz' \delta (z-z') (N- {1\over N}) M(S)  $$

\noindent which simply gives the process of a string remaining the same string.
The
fourth term of eq(A.1) has similar contribution as the third .Adding
all these contributions one arrives at eq.(19).

  Now, the next term in the definition of $d_{M}(S)$,  similarly can be
  obtained as

$$ {\delta^2 M(S)\over {\delta \phi(z)\delta \phi(z')}} = 2\int dz \int
dz' \delta (z-z') \delta (z-x)\delta (z-y)U(x,y)_{bb}U(y,x)_{aa}
\eqno(A.5)  $$

\noindent This refers to a physical process of the string splitting to two
loops
(Fig.2a).

  We now look at the coefficient $D_M(S_1,S_2)$. One obtains,

$${\delta M(S_1)\over {\delta A^a_i(z)}}{\delta M(S_2)\over {\delta A^a_i(z)}}
$$$$ = \bigg[\int^{x_1}_{y_1}dz \phi(x_1)_{\alpha\beta}[T_a(z)U(x_1,y_1)]
_{\beta\gamma}\phi(y_1)_{\gamma\sigma}U(y_1,x_1)_{\sigma\alpha}$$$$
+\int^{y_1}_{x_1}dz \phi(x_1)_{\alpha\beta}U(x_1,y_1)_{\beta\gamma}
\phi(y_1)_{\gamma\sigma}[T_a(z)U(y_1,x_1)]_{\sigma\alpha}\bigg]$$$$
\bigg[ \int^{x_2}_{y_2} dz'\delta (z-z')\phi(x_2)_{pq}[T_a(z')
U(x_2,y_2)]_{qr}\phi(y_2)_{rs}U(y_2,x_2)_{sp}$$$$
+ \int^{y_2}_{x_2} dz' \delta (z-z') \phi(x_2)_{pq}U(x_2,y_2)
_{qr}\phi(y_2)_{rs}[T_a(z')U(y_2,x_2)]_{sp}\bigg]\eqno(A.6)$$

Using eq.(8) and with usual manipulations the first term in the above
product takes the form;

$$ \int^{x_1}_{y_1} dz\int^{x_2}_{y_2}dz'\delta(z-z') \bigg[\phi(x_1)_
{\alpha\beta}U(x_1,z)_{\beta j}U(z',y_2)_{jr}\phi(y_2)_{rs}$$$$
U(y_2,x_2)_{sp}\phi(x_2)_{pq}U(x_2,z')_{qi}U(z,y_1)_{i\gamma}\phi(y_1)_
{\gamma\sigma}U(y_1,x_1)_{\sigma\alpha} $$$$ -{1\over N}\bigg\{ \phi(x_1)_
{\alpha\beta}U(x_1,z)_{\beta\sigma}U(z,y_1)_{\sigma\gamma}\phi(y_1)_
{\gamma\delta}U(y_1,x_1)_{\delta\alpha}$$$$ \phi(x_2)_{pq}U(x_2,z')_
{qj}U(z',y_2)_{jr}\phi(y_2)_{rs}U(y_2,x_2)_{sp}\bigg\}\bigg]
\eqno(A.7)   $$

\noindent The above expression describes a process where two strings join
together
to produce one string with four matter fields S(4) (Fig.3a) and the
two strings remaining unchanged with a suppression factor of 1/N.The
rest of the three terms of eq.(A.6) give rise to similar physical
processes.

  Exactly analogous computation for ${\delta M(S_1)\over {\delta\phi(z)}}
{\delta M(s_2)\over {\delta\phi(z)}}$ leads to fusion of two strings to
produce one string as shown in Fig.3b.

  Finally the term $D_{WM}(L,S)$ can be expressed as the sum of two terms like,

$$ \oint_L dz'\int dz'' \delta(z'-z'') \bigg[U(x',y')_{a\gamma}
\phi(y')_{\gamma\delta}U(y',x'')_{\delta\alpha}\phi(x')_{\alpha a}
$$$$ -{1\over N}\bigg\{ U(x',y')_{ab}U(y',z')_{bd}U(z',x')_{da}
\phi(x'')_{\alpha\beta}$$$$U(x'',z'')_{\beta n}U(z'',y'')_{n\gamma}
\phi(y'')_{\gamma\delta}U(y'',x'')_{\delta\alpha}\bigg\}
\bigg]   $$

\noindent leading to interpretation in terms of physical processes as shown in
Fig.4.

 \vfill
 \eject

\centerline {\bf { REFERENCES:}}

\item{1.}   S. Mandelstam, Phys.\ Rev., {\bf 175}, 1580 (1968); Phys.\ Rev.,
             {\bf D19}, 2391 (1979).

\item{2.}   A. M. Polyakov, Phys.\ Lett., {\bf B82}, 247 (1979); Nucl.\
            Phys., {\bf B164}, 171 (1980).

\item{3.}   Yu. Makeenko and A. A. Migdal, Phys.\ Lett., {\bf B88} 135,
            (1979); Nucl.\ Phys., {\bf B188}, 269 (1981).

\item{4.}   G. 't Hooft, Nucl.\ Phys., {\bf B72}, 461 (1974).

\item{5.}   V. Kazakov, Mod.\ Phys.\ Lett., {\bf A4}, 2125 (1989);
            A. A. Migdal Phys.\ Rep., {\bf C102}, 199 (1985);
            T  Banks, M. Douglas, N. Seiberg and S. Shenker, Phys.\ Lett.,
            {\bf B238}, 279 (1990);
            D. J. Gross and A. A. Migdal, Nucl.\ Phys.,{\bf B340}, 333 (1990);
            R. Dijgraaf, H. Verlinde and E. Verlinde, Nucl.\ Phys., {\bf B348}
            , 435 (1991);
            F. David, Mod.\ Phys.\ Lett.,{\bf A5}, 1019 (1990);
            M. Fukuma, H. Kawai and T. Nakayama, Int.\ J.\ Phys., {\bf A6},
            1385 (1991).

\item{6.}   S. Dalley , Mod.\ Phys.\ Lett.,{\bf A7}, 1651 (1992);
            Phys.\ Lett., {\bf B298}, 79 (1993); Phys.\ Rev.,
            {\bf D47}, 2517 (1993).

\item{7.}   B. Sakita, Phys.\ Rev.,{\bf D21}, 1067 (1980).
            A. Jevicki and B. Sakita, Nucl.\ Phys., {\bf B165}, 511 (1980);
            {\bf B185}, 89 (1981); Phys.\ Rev., {\bf D22}, 467 (1980).

\item{8.}   J. Maharana and L. P. Singh, Phys.\ Lett.,{\bf B142}, 177 (1984);
            Phys.\ Rev., {\bf D31}, 3162 (1984).

\item{9.}   A. J. Hanson, T. Regge and C.Titelboim,{\sl Costrained Hamiltonian
            systems} (Academica Nationale dei Lincei,Roma,1976);E. C. G.
            Sudarshan and N. Mukunda, {\sl Classical dynamics; a modern
            perspective} (Wiley, New York, 1974).

\item{10.}  J. L. Gervais and A. Jevicki, Nucl.\ Phys.,{\bf B110}, 93 (1976).

\vfill
\eject
\centerline {\bf FIGURE CAPTION}

\bigskip

Figure $1a$. Depicts a loop  splitting into two loops due to the
operator $d_{W}(L)$.

Figure $1b$. Describes the process where a loop annihilates with its
conjugate.

Figure $1c$. Shows fusion of two loops to form a single loop. The
term $D_{W}(L_1,L_2)$ is responsible for the processes of figure $1b$
and $1c$.

Figure $2a$. Depicts splitting of a closed string into two loops.

Figure $2b$. Shows a string splitting to form a loop and a string.

Figure $2c$. Describes splitting of a string to two strings.

Figure $3a$. Is the fusion of two strings to form another string with
four matter fields.

Figure $3b$. Depicts fusion of two strings to form a string in the
configuration of the figure of eight.

Figure $4$. Shows how a string and a loop fuse to form another string.
\vfill
\eject
\end